\documentclass[aps,prl,reprint,superscriptaddress]{revtex4-1}

\usepackage{graphicx}

\newcommand{\width}{\Lambda}
\newcommand{\fracDelta}{\Delta^{\prime}}
\newcommand{\Fexp}{\mu}

\begin{document}

%\preprint{}

\title{Viscoelastic scaling regimes for marginally--rigid fractal spring networks}


\author{David Head}
\email[]{d.head@leeds.ac.uk}
\affiliation{School of Computing, University of Leeds, Leeds LS2 9JT, United Kingdom}
%\thanks{}


\date{\today}

\begin{abstract}
	A family of marginally--rigid (isostatic) spring networks with fractal structure up to a controllable length was devised and the viscoelastic spectra $G^{*}(\omega)$ calculated. Two non--trivial scaling regimes were observed, (i)~$G^{\prime}\approx G^{\prime\prime}\propto\omega^{\Delta}$ at low frequencies, consistent with $\Delta=1/2$; (ii)~$G^{\prime}\propto G^{\prime\prime}\propto\omega^{\fracDelta}$ for intermediate frequencies corresponding to fractal structure, consistent with a theoretical prediction $\fracDelta=(\ln3-\ln2)/(\ln3+\ln2)$. The cross--over between these two regimes occurred at lower frequencies for larger fractals in a manner suggesting diffusive--like dispersion. Solid gels generated by introducing internal stresses exhibited similar behaviour above a low--frequency cut--off, indicating the relevance of these findings to real--world applications.
\end{abstract}

\pacs{AAA}

%\keywords{}

\maketitle

%
% Introduction: Motivation, context, summary of main findings.
%
{\em Introduction.}---Many soft matter and complex systems exhibit power law rheology over a broad frequency range, manifested as parallel scaling of the linear storage and loss moduli $G^{\prime}(\omega)\propto G^{\prime\prime}(\omega)\propto\omega^{\Delta}$~\cite{Aime2018,Hang2021,Rathinaraj2021}, or equivalently a power--law relaxation spectrum~\cite{Martin1991,Zaccone2014,Rizzi2020}. Relating this scaling to the underlying causal mechanisms would guide the selection of synthesis pathways producing desirable material properties in a number of application domains~\cite{Hung2015,HeurtaLopez2021}, but is not yet generally possible. Of the many potential contributions, slow structural relaxation~\cite{Fielding1999,Kroy2007,Mulla2019} cannot be a prerequisite, as $\Delta\ll1$ has been observed in protein hydrogels with permanent crosslinks and no unfolding~\cite{AufderhorstRoberts2020,Hughes2021,AufderhorstRoberts2022}. There must therefore be processes capable of generating broad distributions of relaxation times that do not require topological changes to material microstructure.

It has been hypothesised that the broad distribution of relaxation times derives from a similarly broad distribution of structural length scales~\cite{Rathinaraj2021,Hughes2021}. Such structure emerges naturally from cluster aggregation processes, which can produce a scale--invariant, or fractal, geometry up to a characteristic maximum length~\cite{Meakin1992,Jungblut2019,Hanson2020}. Calculations for  branched fractal polymers predict power--law rheology with a $\Delta$ that depends on the fractal dimension $d_{\rm f}$ and solvent condition~\cite{Muthukumar1985,Martin1989}, but cannot explain values $\Delta\ll1$ without invoking unphysical fractal dimensions~\cite{Muthukumar1989}. A quite different mechanism applies to tenuous solids close to their rigidity transition, defined here as when $G^{\prime}(\omega=0)$ first becomes non-zero, such as at gelation. Normal mode analysis of athermal elastic packings have demonstrated that the lowest eigenvalue, and hence relaxation frequency, vanishes as the rigidity transition is approached, resulting in an arbitrarily broad relaxation spectrum~\cite{Lemaitre2006,Silbert2009,Huisman2011,Milkus2017}.

The relative contributions of these two non-exclusive mechanisms to power--law viscoelasticity can be elucidated by the construction and analysis of model systems that are both fractal and marginally rigid. Such systems are considered here. A family of athermal spring networks based on the Sierpinski triangle was devised in which every node connects to $z=2d=4$ others, equalling the isostatic threshold when frames first become rigid~\cite{Calladine1978}. These correlated networks~\cite{Michel2019,Zhang2019} are fractal up to a controllable length, as in aggregation--derived structures~\cite{Meakin1992,Jungblut2019,Hanson2020}, and the lower limit of this length produces the kagome lattice~\cite{Sun2012,Mao2013}. These networks are related to those of Machlus {\em et al.}~\cite{Machlus2021} that however are not locally isostatic everywhere. A matrix--based solver was then used to estimate $G^{*}(\omega)$ over a broad range of $\omega$, and two non--trivial scaling regimes found. Low frequencies were consistent with $\Delta=\frac{1}{2}$, also measured for bond--diluted networks, and attributed to marginal rigidity. This matches the exponent for crosslinker--unbinding in semi\-flexible polymer networks~\cite{Broedersz2010} but has a distinct origin. It also matches the intermediate scaling regime for thermal Rouse modes in linear polymers, extended to fractal branched polymers by Muthumukar~\cite{Muthukumar1985,Muthukumar1989}, but again a causal relation seems improbable. Conversely, an intermediate frequency regime exhibited $\Delta<\frac{1}{2}$ that was related to the spectral dimension of Sierpinski fractals~\cite{Liu1984}. The crossover frequency between the two regimes varied with the maximum fractal length in a manner suggesting diffusive--like dispersion. The generalisation of these findings to arbitrary fractals are discussed at the end.

\begin{figure}[htbp]
	\centerline{
		\includegraphics[width=9.5cm]{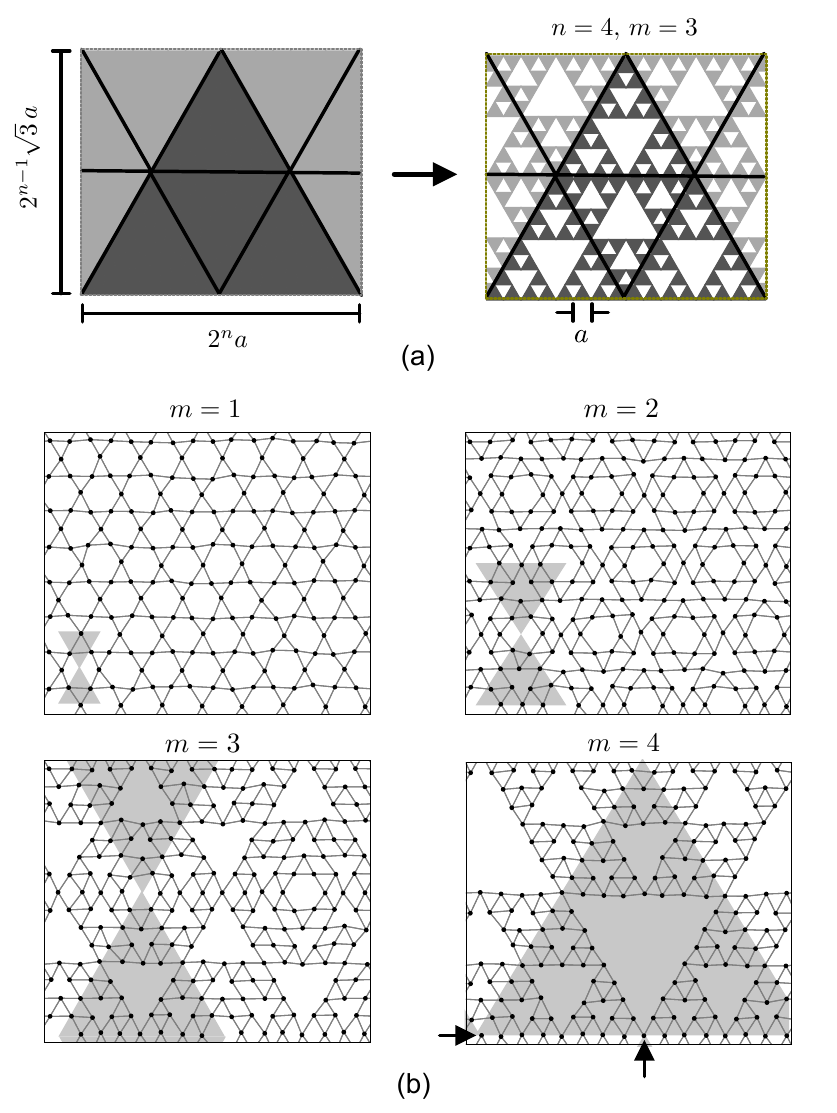}
	}
	\caption{(a) Schematic of network generation for system size $2^{n}\times 2^{n}$ and maximum fractal length $2^{m}$. Upward (dark) and downward (light) oriented triangles spanning a periodic box were subdivided into 4 sub-triangles (left). Each sub-triangle was then recursively subdivided into 3 triangles each over $m=3$ further iterations, as per standard Sierpinksi triangles (right). (b) Examples of networks for $n=4$ and $1\leq m\leq4$. Dots denote nodes and straight lines denote springs. The maximum fractal extent is indicated by the shaded triangles and the origin has been shifted so all nodes are clearly visible. Arrows indicate the two nodes for $m=n$ with $z=2$. Images for $n=6$ are provided in~\cite{suppInf}.}
	\label{f:methods}
\end{figure}

%
% Methods.
%
{\em Methods.}---Arrays of $2^{n}\times 2^{n}$ nodes were assembled onto regular triangular lattices with spacing $a$ in a rectangular box of dimensions commensurate to the lattice. The full system was partitioned into two system-spanning triangles with opposite orientations, each with $2^{n}$ nodes along each edge. Each triangle was then subdivided into 4 equal-sized sub-triangles with edge length $2^{n-1}$, and so on recursively, generating sub-triangles with edge lengths $2^{n-2}$, $2^{n-3}$ {\em etc.}, for $n-m$ iterations. For the remaining $m$ iterations, only the 3 sub-triangles at vertices were recursed as per standard Sierpinski triangle generation, thus generating fractal structure for lengths $2^{m}a$ down; see Fig.~\ref{f:methods}(a). Edges of the smallest triangles after the final $n^{\rm th}$--iteration were mapped onto Hookean springs, excluding those lying along the edges of the major triangles with $2^{m}$ nodes along each edge. Isolated nodes with no attached springs were removed. To remove co\-linear springs, $x$ and $y$--coordinates of all nodes were perturbed by small Gaussian displacements with mean zero and variance $(\sigma_{j}a)^{2}$, where $\sigma_{j}=0.05$. The natural lengths of all springs were set to the inter-node separation after this perturbation, so there were no internal stresses. When internal stresses were required, all natural spring lengths were additionally changed by a Gaussian random variable with zero mean and variance $(\width a)^{2}$, and nodes non-linearly moved to coordinates obeying static equilibrium using FIRE~\cite{Bitzek2006}.

Elastic forces on network nodes were required to balance drag forces due to the surrounding fluid throughout cycles of simple oscillatory shear at frequency $\omega$. Hydrodynamic interactions~\cite{Dennison2016,Head2019} were absent, therefore the only degrees of freedom were the complex displacement 2--vectors for each node $\alpha$, written ${\bf u}^{\alpha}(t)={\bf u}^{\alpha}_{\omega}e^{i\omega t}$ in terms of the complex amplitudes ${\bf u}^{\alpha}_{\omega}$ (real part understood). The drag force on node $\alpha$ was $\zeta\left[{\bf v}^{\alpha,\rm aff}(t)-\partial_{t}{\bf u}^{\alpha}(t)\right]$ in terms of the drag coefficient $\zeta$ and the affine fluid velocity at the position $(x^{\alpha},y^{\alpha})$ of node $\alpha$, ${\bf v}^{\alpha,{\rm aff}}(t)=(\gamma(t)y^{\alpha},0)$~\cite{Yucht2013}. After cancelling all factors of $e^{i\omega t}$, the force balance equations between drag (left--hand side) and elastic  (right--hand side) forces was
\begin{equation}
	\zeta\left(
		{\bf v}^{\alpha,\rm aff}_{\omega} - i \omega {\bf u}^{\alpha}_{\omega}
	\right)	
	=
	\sum_{\beta\sim\alpha}
	H^{\alpha\beta}\left(
		{\bf u}^{\beta}_{\omega}
		-
		{\bf u}^{\alpha}_{\omega}
	\right)
	,
	\label{e:forceBalance}
\end{equation}
with $H^{\alpha\beta}$ the $2\times 2$ Hessian matrix for a single spring of stiffness $k$ between connected nodes $\alpha$ and $\beta$,
\begin{equation}
  H^{\alpha\beta}_{ij}
  =
  k\,
  \hat{t}^{\alpha\beta}_{i}
  \hat{t}^{\alpha\beta}_{j}
  +
  \frac{\tau^{\alpha\beta}}{\ell^{\alpha\beta}}
  \left(
    \delta_{ij}
    -
    \hat{t}^{\alpha\beta}_{i}
    \hat{t}^{\alpha\beta}_{j}    
  \right)
  \label{e:localHessian}
\end{equation}
in terms of the unit vector $\hat{\bf t}^{\alpha\beta}$ from $\alpha$ to $\beta$, the inter-node separation $\ell^{\alpha\beta}$, and the spring tension $\tau^{\alpha\beta}$ with vanishes in the absence of internal stresses. Equations (\ref{e:forceBalance}) were assembled into a global solution vector of all nodal complex amplitudes, and the resulting matrix equation, including Lees--Edwards shifts across sheared boundaries~\cite{Allen1987}, solved using the SuperLU sparse direct method~\cite{Demmel2003} as described previously~\cite{Head2019}.

%
% Isostatic fractal networks without prestress.
%
{\em Viscoelastic spectra.}---Example networks are presented in Fig.~\ref{f:methods}(b). Every node was connected to $z=4$ others, except for $m=n$, when 2 nodes in the entire system had $z=2$, resulting from the largest fractal triangles intersecting tip--to--base through the periodic boundaries, rather than tip--to--tip as for $m<n$. The kagome lattice~\cite{Sun2012,Mao2013} corresponds to $m=1$, and as $m\rightarrow n$, Sierpinksi triangles of fractal dimension $d_{\rm f}=\ln 3/\ln 2$~\cite{FalconerBook} became evident. As derived in supplementary materials~\cite{suppInf}, the total number of springs  $N_{\rm spring}=6(3^{m}-2^{m})4^{n-m}$ and nodes $N_{\rm node}=3(3^{m}-2^{m})4^{n-m}+\delta_{nm}$, with $\delta_{nm}$ the Kronecker delta. Thus the mean coordination number $\langle z\rangle=2N_{\rm bond}/N_{\rm node}=4$ for $m<n$, with a small correction $4-\langle z\rangle\sim4\cdot3^{-(m+1)}$ for $m=n$. The pebble game method~\cite{Jacobs1995} confirmed all nodes belonged to a single rigid cluster, and there were no redundant springs -- that is, springs that can be removed without loss of rigidity -- except for a trivial ${\mathcal O}(1)$ set deriving from rigid-body motion of the whole network. This means that all springs become stressed, with either positive or negative tension, when the network is sheared.
% All data presented here has been confirmed to have converged with system size, which primarily affected low--$\omega$ data; see Fig.~S2 in~~\cite{suppInf}.

\begin{figure}[htbp]
	\centerline{
		\includegraphics[width=9.5cm]{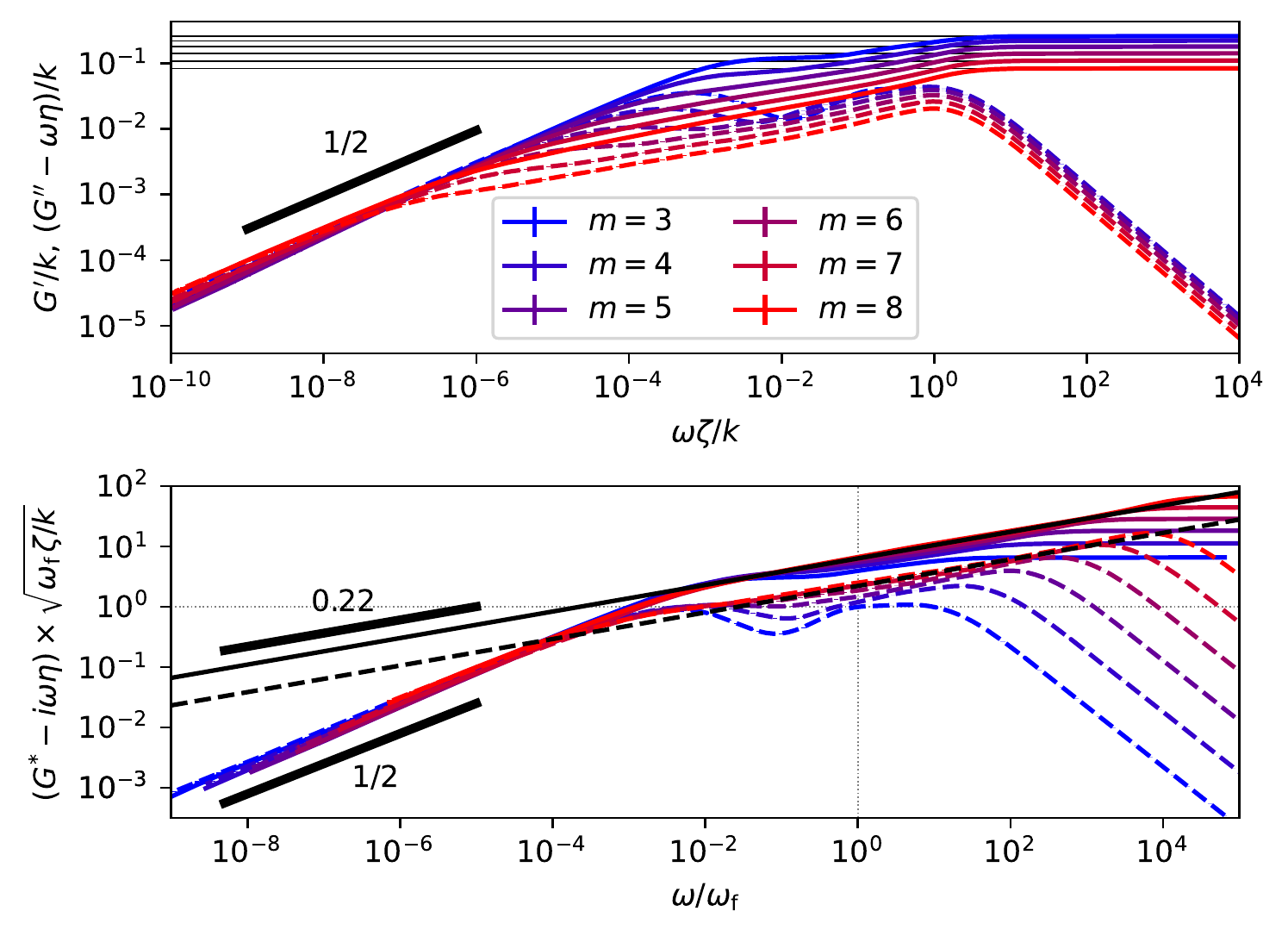}
	}
	\caption{Viscoelastic spectra for networks of $2^{n}\times2^{n}$ nodes with $n=10$, with fractal structure up to a length $\propto2^{m}$ with $m$ given in the legend. Solid lines denote $G^{\prime}(\omega)$ scaled to the spring constant $k$, and dashed lines denote the network contribution ({\em i.e.} less the solvent contribution $\eta\omega$ for viscosity $\eta$) to $G^{\prime\prime}(\omega)$. Thin horizontal lines give the affine prediction, $G^{\prime}_{\rm aff}=\frac{\sqrt{3}}{2}k\frac{3^{m}-2^{m}}{4^{m}}$. The same data is replotted against $\omega/\omega_{\rm f}$ with $\omega_{\rm f}\zeta/k=(2^{m})^{-2}$ in the lower panel, with fits to the intermediate scaling regime shown. In both panels, thick line segments have the annotated slope.}
	\label{f:isoCollapse}
\end{figure}

Viscoelastic spectra $G^{*}(\omega)$ for different $m$ are given in Fig.~\ref{f:isoCollapse}.
%Affine response, in which each spring deforms as per the global strain, was expected for high frequencies.
It is straightforward to derive the affine prediction $G^{'}_{\rm aff}/k=\frac{\sqrt{3}}{2}\left(\frac{3^{m}-2^{m}}{4^{m}}\right)$~\cite{suppInf}, which matches the numerical results for large~$\omega$, confirming affinity at frequencies above the highest one-spring mode~\cite{Huisman2010}. By contrast, for low frequencies a power--law scaling $G^{\prime}(\omega)\approx G^{\prime\prime}(\omega)\propto\omega^{\Delta}$ with $\Delta\approx0.5$ was observed. This is consistent with the Kramers--Kronig relation specialized to power--law $G^{*}(\omega)$, which requires $G^{\prime\prime}/G^{\prime}=\tan(\Delta\pi/2)$~\cite{Chambon1987}. The exponent $\Delta=\frac{1}{2}$ has been predicted by effective medium~\cite{Yucht2013} and scaling~\cite{Tighe2012} theories for non--fractal systems.
%, and observed in PDMS experiments at the gelation point~\cite{Winter1986}. 
% [below dropped as is now covered in the introduction]
% Power--law scaling with exponents $\geq\frac{1}{2}$ have been predicted at the gelation transition for polymeric systems, with the exponent depending on gel fractal dimension and solvent conditions~\cite{Muthukumar1989,Martin1989,Martin1991}. However, these calculations only consider relaxation processes related to mass diffusion, whereas for the model here, stress relaxation can only be due to overdamped network elastic modes.
As the maximum fractal length $\propto2^{m}$ was increased, an intermediate frequency regime emerged in which both $G^{\prime}(\omega)$ and $G^{\prime\prime}(\omega)$ scaled as a power law $\approx\omega^{\fracDelta}$ with $\fracDelta\approx0.22$. The fitted ratio $G^{\prime\prime}/G^{\prime}\approx0.35$ was consistent with the  Kramers--Kronig previously mentioned, {\em i.e.} $\tan(\fracDelta\pi/2)\approx\tan(0.22\,\pi/2)\approx0.36$, suggesting this scaling will persist to $\omega\rightarrow0$ for arbitrarily large fractals $2^{m}\rightarrow\infty$. Furthermore, it was possible to collapse curves for $m\geq 3$ onto a single master curve for low and intermediate frequencies by scaling $\omega$ by $\omega_{\rm f}\propto(2^{m})^{-2}$, and $G^{*}(\omega)$ by $(\omega_{\rm f})^{1/2}$ so as to preserve $G^{\prime}\approx G^{\prime\prime}$ for low $\omega$. Since the maximum fractal length is $\propto 2^{m}$, this collapse suggests $\omega_{\rm f}\sim q^{2}$ for the wavelength $\propto q^{-1}\propto 2^{m}a$, which is a diffusive-like dispersion relation~\cite{Aime2018}. The same collapse was also observed for $m=1,2$ but to a different master curve for reasons that are not yet understood; see Fig.~S1 of~\cite{suppInf}. The scaling of the regimes, and the width of the crossovers between them, do not depend on system size as shown in Fig.~S2 of~\cite{suppInf}.

% Viscoelastic spectra, unscaled and then scaled.

$G^{*}(\omega)$ was also calculated for networks generated by random bond dilution, where springs are present with a probability $p$. Such networks are known to exhibit a rigidity transition at a critical dilution $p=p_{\rm c}$ that is well--defined for infinite systems~\cite{SahimiBook}. For $p$ sufficiently close to $p_{\rm c}$, it was found that the viscoelastic spectra were again consistent with $G^{\prime}(\omega)=G^{\prime\prime}(\omega)\propto\omega^{1/2}$, as demonstrated in Fig.~S3 of~\cite{suppInf}. The power--law regime with an exponent $\approx0.4$ previously reported for the lowest frequencies attained by more general numerical schemes~\cite{Yucht2013,Dennison2016} is also apparent in this figure, and identified here as an intermediate frequency regime.
%The scaling did not sensitively depend on the existence of `dangling' regions not rigidly connected to the stress--bearing backbone, as identified by the pebble game method.

The exponents for the viscoelastic scaling regimes were confirmed by the scaling framework of Tighe, which, after eliminating a correlation length, predicts $G^{\prime}\propto G^{\prime\prime}\propto\omega^{1-\nu}$, with $\nu$ related to the variation of the magnitude of non--affine displacements, $NA\propto\sum_{\alpha}|{\bf u}^{\alpha}-{\bf u}^{\alpha,\rm aff}|^{2}\propto\omega^{-\nu}$, where ${\bf u}^{\alpha,\rm aff}$ is the affine displacement for node $\alpha$~\cite{Tighe2012}. For low frequencies, NA decayed with an exponent $\nu\approx0.5$ as shown in Fig. S4 of~\cite{suppInf}, consistent with theoretical considerations~\cite{Wyart2008} and the viscoelasticity scaling $\Delta=1-\nu=\frac{1}{2}$. For the intermediate frequency regimes, the prediction became $\nu=1-\fracDelta$, and thus $\nu\approx1-0.22=0.78$ for fractal networks and $\approx1-0.4=0.6$ for bond--diluted networks, which is again consistent with the NA data in the same figure.

%
% Liu theory and prediction of fractal scaling exponent.
%
{\em Derivation of $\fracDelta$.}---Liu recursively generated the dynamical matrix for isolated Sierpinski spring networks to derive the scaling of the density of states~\cite{Liu1984}. The same approach can be extended to derive a prediction for the slope $\fracDelta$ of the intermediate regime. Let $F(\tau)$ denote the contribution to rigidity by processes with relaxation time~$\tau$, and assume a power--law tail $F(\tau)\sim F_{0}\tau^{-\Fexp}$.  Following Liu~\cite{Liu1984}, for each additional level of recursion, both the diagonal dynamical matrix elements and the number of degrees of freedom increase by a factor of 3, whereas the effective stiffness halves. For over\-damped systems as here, the diagonal scaling suggests an effective damping coefficient that increases threefold, and hence the relaxation time $\tau$ --- being proportional to damping and inversely proportional to stiffness --- increases by a factor of~6. Using $\tau$ and $\tau^{\prime}$ to denote relaxation times between successive levels of recursion, this means that
\begin{equation}
	\frac{3}{2}F_{0}\left(\tau^{\prime}\right)^{-\Fexp}{\rm d}\tau^{\prime}
	=F_{0}\tau^{-\Fexp}{\rm d}\tau
	\:,
	\label{e:Liulike}
\end{equation}
with the left--hand side factors 3 for the increase in degrees of freedom, and $\frac{1}{2}$ for the reduction in stiffness. This second factor is absent in~\cite{Liu1984}. Inserting $\tau^{\prime}=6\tau$ into (\ref{e:Liulike}) gives $\frac{3}{2}6^{1-\Fexp}=1$, or $\Fexp=2\ln3/\ln6$. Using the relation $F(\tau)\propto\tau^{-\Fexp}$ to $G^{*}(\omega)\propto\omega^{\Fexp-1}$~\cite{Martin1991,Zaccone2014},
\begin{equation}
	\fracDelta
	=
%	\frac{2\ln3}{\ln6}-1
%	=
	\frac{\ln3-\ln2}{\ln3+\ln 2}
	\approx
	0.226\:,
	\label{e:fracScalPred}
\end{equation}
in good agreement with the measured value $\fracDelta\approx0.22$.

\begin{figure}[hbt]
	\centerline{
		\includegraphics[width=9.5cm]{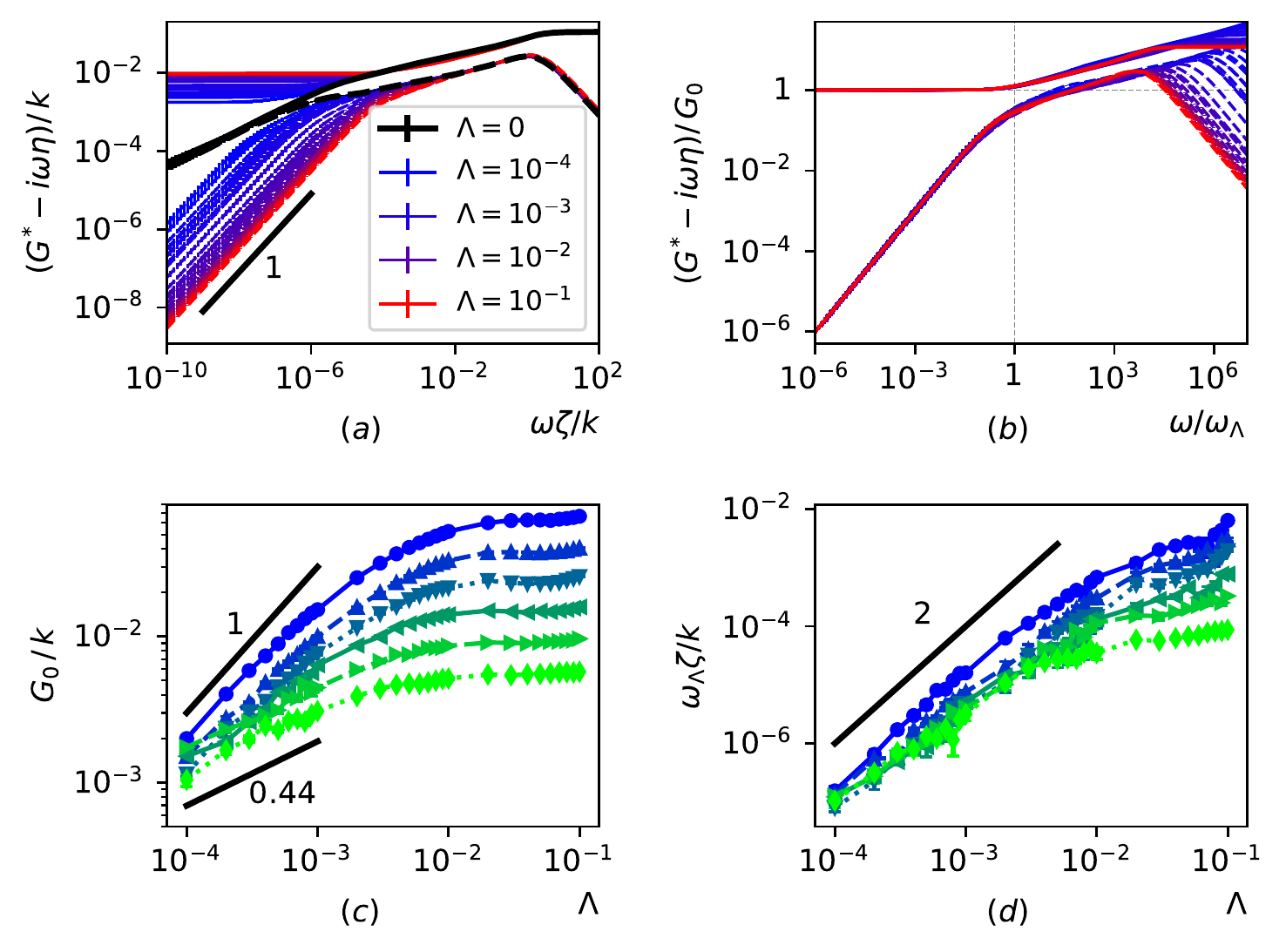}
	}
	\caption{(a)~Viscoelastic spectra for fractal length $\propto2^{m}$ with $m=7$, varying the magnitude of internal stress $\width$. The upper set of curves (solid) are $G^{\prime}$, and the lower (dashed) are the network contribution to $G^{\prime\prime}$. Selected values of $\width$ are given in the legend; curves for other $\width$ interpolate these. (b)~The same data replotted with $G^{*}$ scaled by $G_{0}=G^{\prime}(\omega=0)$ and $\omega$ scaled by $\omega_{\width}$. (c) and (d) give the corresponding values of $G_{0}$ and $\omega_{\width}$ as functions of $\width$, increasing incrementally from $m=3$ (top curves) to $m=8$ (bottom curves). For all figures, the solid line segments have the annotated slopes.}
	\label{f:intStressCollapse}
\end{figure}

{\em Internal stresses.}---Geometries that sustain states of self--stress can be rigid when networks of the same topology, but with geometries that permit fewer or no states of self--stress, are non--rigid~\cite{Calladine1978,Pellegrino1986,Vermeulen2017,Bose2019}. Internal stresses were introduced by changing the natural spring lengths by a random amount $\propto\Lambda a$ as described earlier; an example is given in Fig.~S5 of~\cite{suppInf}. As before, application of the pebble game confirmed the lack of redundant bonds for these perturbed geometries, {\em i.e.} all springs became either stretched or compressed, as evident from the figure.
% Original sentence
%Introducing internal stresses by changing the natural spring lengths by a random amount $\propto\Lambda a$, as described earlier, should therefore produce a solid--like response $G^{\prime}(0)>0$. 
%
Viscoelastic spectra varying $\Lambda$ with $m$ fixed are shown in Fig.~\ref{f:intStressCollapse}(a). The spectra for $\width>0$ match those for $\width=0$ for high frequencies, changing to a solid response, with constant $G^{\prime}(\omega)$ and $G^{\prime\prime}(\omega)\propto\omega$, below a frequency $\omega_{\width}$ that increases with $\width$. Furthermore, it was possible to simultaneously collapse both the low and intermediate frequency regimes for $\width>0$ by scaling $G^{*}$ by the plateau modulus $G_{0}=G^{\prime}(\omega=0)$, and the frequency $\omega_{\width}$ such that $G^{\prime\prime}(\omega^{*})\propto\omega_{\width}$ for the lowest frequencies, as shown in Fig.~\ref{f:intStressCollapse}(b). The $G_{0}$ and $\omega_{\width}$ used to achieve this collapse are given in Figs.~\ref{f:intStressCollapse}(c) and (d) respectively, for a range of maximum fractal lengths $\propto 2^{m}$.

Both the plateau modulus $G_{0}$ and the crossover frequency $\omega_{\width}$ smoothly approach zero as $\width\rightarrow0$, suggesting internal stresses generate rigidity continuously~\cite{Merkel2019} by removing low--frequency response modes. A similar trend was seen for adding random springs, and also for removing springs which induces a crossover to fluid--like (rather than solid--like) response starting at low frequencies, as shown in Fig.~S6 of~\cite{suppInf}. Unlike internal stresses, such perturbations also modify the network connectivity. For $\width\ll1$, the crossover frequency data is consistent with the quadratic variation $\omega_{\width}\propto\width^{2}$ for all~$m$. The variation of $G_{0}$ depends upon whether $\omega_{\width}$ falls in the low or intermediate--frequency regimes; that is, whether $\omega_{\width}<\omega_{\rm f}$ or $\omega_{\width}>\omega_{\rm f}$, which in turn is controlled by the fractal length $\propto 2^{m}$. For small $m$, $\omega_{\width}<\omega_{\rm f}$ for all $\width$ considered, and $G_{0}\propto\width\propto\omega_{\width}^{1/2}$, consistent with a low--frequency cut--off when $G^{*}\propto\omega^{1/2}$ scaling is obeyed. Conversely, for larger $m$ when $\omega_{\width}>\omega_{\rm f}$ becomes accessible, the data is consistent with $G_{0}\propto\width^{2\fracDelta}\propto\omega_{\width}^{\fracDelta}$, which itself is consistent with a frequency cut--off in the intermediate regime $G^{*}\propto\omega^{\fracDelta}$. Qualitatively similar results have been observed for elastic sphere packings under compression and elastic beams under shear~\cite{Wyart2005,Vermeulen2017}, but it is unclear if there is any relationship between the exponents in these systems.

%
% Discussion.
%
{\em Discussion.}---It has been shown that the design of isostatic networks~\cite{Sadjadi2021} fractal up to an arbitrarily--large length is possible, and that for the Sierpinski triangle--based spring networks considered here, power--law viscoelastic scaling was observed with an exponent $\fracDelta$ that can be theoretically derived. That this scaling survives above a cut--off frequency for systems into the solid phase indicates relevance to real--world applications utilising post--gelled materials. However, a general relation between $\fracDelta$ and the fractal dimension $d_{\rm f}$ is not yet available, as existing expressions~\cite{Muthukumar1985,Martin1989} have limited applicability, and the arguments of Liu~\cite{Liu1984} cannot be easily generalised to arbitrary $d_{\rm f}$. The crossover frequency $\omega_{\rm f}$ is not related a Boson--like peak as this vanishes at isostaticity~\cite{Silbert2009,Wyart2005}; normal mode analysis might help identify its role in mode propagation.
%Liu derived an expression for Sierpinski fractals in higher dimensions~\cite{Liu1984} which, using the same argument as before, predicts $\fracDelta=(d_{\rm f}-1)/(d_{\rm f}+1)$. Although valid for the Sierpinski triangle--based networks here, it predicts $\Delta^{\prime}\approx0.30$ for bond--diluted networks with a backbone dimension $d_{\rm f}\approx1.86$~\cite{Jacobs1995}, significantly different from the measured value $\approx0.4$. It also does not predict $\Delta=\frac{1}{2}$ for non--fractal materials with $d_{\rm f}=d=2$. Thus, a more general relation is needed.
Experimental validation of these trends should be possible by controlling the size of the fractal mesostructure (measured {\em via} scattering) varying the volume fraction and/or the reaction rate~\cite{AufderhorstRoberts2020,Hughes2021}. A broad frequency range at gelation would be accessible using time--cure superposition~\cite{Adolf1989}, and should reveal an intermediate power--law regime that is here predicted to extend to lower frequencies for larger fractal lengths. Quantitative agreement would however require the development of 3--dimensional models with realistic aggregation kinetics.
In addition, any future experimental validation will require quantitative predictions for the cross--over frequency between low and intermediate scaling regimes, necessitating 3--dimensional modelling. Further work investigating a broader range of fractal structure with $d=3$, including dynamically--generated stochastic fractals as opposed to the deterministic fractals considered here, would help alleviate these challenges and improve our understanding of the link between fractal structure and viscoelastic response for this important class of materials.

%
% Acknowledgements.
%

\begin{acknowledgments}
	The author would like to thank Wouter Ellenbroek, Xiaoming Mao, Anders Aufderhorst--Roberts and Benjamin Hanson for discussions.
\end{acknowledgments}

\bibliography{fractalGel}

% Alternatively ...
%\begin{thebibliography}{99}
%\bibitem{Bray2001a} {\em Cell Movements}, D. Bray (Garland, New York, 2001).
%\end{thebibliography}

\end{document}